\documentclass[aps,prl,amsmath,twocolumn,superscriptaddress,showpacs,floatfix]{revtex4}

\def\PrOS{PrOs$_4$Sb$_{12}$}

\def\LaOS{LaOs$_4$Sb$_{12}$} 
\def\LPOS{La$_{1-x}$Pr$_x$Os$_4$Sb$_{12}$}

\newif\ifpdf
\ifx\pdfoutput\undefined
\pdffalse 
\else
\pdfoutput=1 
\pdftrue
\fi

\ifpdf
\usepackage[pdftex]{graphicx}
\else
\usepackage{graphicx}
\fi

\begin{document}

\title{The Crystal Field Potential of PrOs$_4$Sb$_{12}$: Consequences for Superconductivity}
\date{\today}
\author{E. A. Goremychkin}
\author{R. Osborn}
\email{ROsborn@anl.gov}
\affiliation{Argonne National Laboratory, Argonne, IL 60439}
\author{E. D. Bauer}
\altaffiliation[Present Address: ]{Los Alamos National Laboratory, Los Alamos, NM 87544, USA}
\author{M. B. Maple}
\author{N. A. Frederick}
\author{W. M. Yuhasz}
\affiliation{Department of Physics and Institute for Pure and Applied Physical Sciences, University of California, San Diego, La Jolla,  CA 92093}
\author{F. M. Woodward}
\author{J. W. Lynn}
\affiliation{NIST Center for Neutron Research, NIST, Gaithersburg, MD 20899-8562}

\ifpdf
\DeclareGraphicsExtensions{.pdf, .jpg, .tif}
\else
\DeclareGraphicsExtensions{.eps, .jpg}
\fi

\begin{abstract}
The results of inelastic neutron scattering provide a solution for the crystal field level scheme in \PrOS, in which the ground state in the cubic crystal field potential of T$_h$ symmetry is a $\Gamma_1$ singlet. The conduction electron mass enhancement is consistent with inelastic exchange scattering, and we  propose that inelastic quadrupolar, or aspherical Coulomb, scattering is responsible for enhancing the superconducting transition temperature. \PrOS\ appears to be the first compound in which aspherical Coulomb scattering is strong enough to overcome magnetic pair-breaking and \textit{increase} T$_c$.
\end{abstract}


\maketitle

Praseodymium filled skutterudite compounds, with general formula PrT$_4$M$_{12}$, where T is one of the transition metals Fe, Ru, or Os, and M is a pnictogen (P, As, or Sb), show a remarkable variety of interesting physical phenomena, including metal-insulator transitions~\cite{Sekine:1997}, quadrupolar heavy fermion behavior~\cite{Aoki:2002,Sugawara:2002}, and superconductivity~\cite{Takeda:2000}.  In particular, \PrOS\ has attracted attention as a heavy fermion superconductor, in which quadrupolar fluctuations may play an important role in the pairing mechanism~\cite{Bauer:2002}.  This proposal is based on two key observations; (i) there is a significant f-electron-induced mass enhancement of the conduction electrons, observed in specific heat, upper critical field~\cite{Bauer:2002,Maple:2002}, and de Haas-van Alphen measurements~\cite{Sugawara:2002b}, and (ii) the magnetic susceptibility indicates that the crystal field ground state is non-magnetic~\cite{Bauer:2002}.

A knowledge of the crystal field ground state is essential to understanding the role of the f-electrons in the superconductivity.  In analyzing the magnetic susceptibility and specific heat, Bauer \textit{et al} considered two possible crystal field models~\cite{Bauer:2002}.  In cubic symmetry, the Pr$^{3+}$ ion splits into a singlet ($\Gamma_1$), a non-magnetic doublet ($\Gamma_3$), and two magnetic triplets ($\Gamma_4$ and $\Gamma_5$).  Crystal field models with either the $\Gamma_1$ singlet or $\Gamma_3$ doublet as ground state were both broadly consistent with the data; in both cases, the $\Gamma_5$ triplet was the lowest excited level estimated to be at less than 1 meV in energy.  A $\Gamma_3$ non-Kramers doublet ground state is of particular interest as it provides the necessary conditions for quadrupolar Kondo fluctuations to be responsible for the heavy fermion behavior~\cite{Cox:1998}, and was favored by analyses of the entropy~\cite{Maple:2003b,Vollmer:2003}.  However, the alternative $\Gamma_1$ singlet ground state has also been proposed following experiments that have explored the crossover to a field-induced ordered phase~\cite{Aoki:2003,Tayama:2003,Rotundu:2004}, so this important question remains unresolved.

Inelastic neutron scattering is the most direct method of determining the crystal field potential and level scheme of metallic rare earth systems. In this report, we present the results of a comprehensive set of measurements of crystal field transitions in \PrOS\ as a function of temperature.  From a simultaneous profile refinement of all the spectra, normalized on an absolute intensity scale, we have concluded that the $\Gamma_1$ singlet is the ground state level. Discrepancies with earlier neutron scattering reports~\cite{Maple:2002} are explained by the need for extra terms in the cubic crystal field Hamiltonian that are required by the $T_h$ point group symmetry~\cite{Takegahara:2001}.  Although our data are inconsistent with a quadrupolar Kondo scenario, we conclude that inelastic quadrupolar fluctuations do play a vital role in enhancing the superconducting transition temperature, through aspherical Coulomb scattering of the conduction electrons~\cite{Fulde:1970}.

We performed our experiments on the same polycrystalline sample that was used for previous neutron scattering measurements; details of its preparation and characterization can be found in Ref. \cite{Maple:2002}. The inelastic neutron scattering experiments have been performed on the time-of-flight Fermi chopper spectrometer LRMECS at the pulsed spallation neutron source IPNS (Argonne National Laboratory, Argonne), and the cold-source triple-axis spectrometer SPINS at the NIST Center for Neutron Research.  The time-of-flight measurements used incident energies of 6, 25, 35, and 60 meV, and were normalized on an absolute intensity scale using a vanadium standard.  The higher energy runs showed no evidence of crystal field excitations above 20 meV.  LRMECS has continuous detector coverage from $2.4^\circ$ to $117.6^\circ$.  The data at the highest angles ($>100^\circ$) are dominated by phonon scattering, but extrapolation of the momentum transfer dependence to low angle shows that the phonon contribution below $30^\circ$ is sufficiently small at all incident energies to be neglected in our analysis.  The SPINS data, which were collected with a fixed final energy of 3.7 meV using a cold BeO filter, showed evidence of a well-resolved transition at 0.7 meV, which decreases in intensity with increasing temperature, \textit{i.e.}, it is a crystal field transition from the ground state.

\begin{figure}[htbp]
\begin{center}
\vspace{-0.1in}

\centerline {
\includegraphics[width=3in]{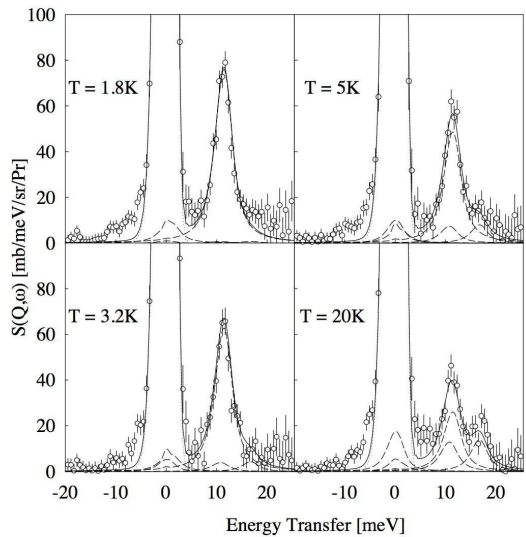}
}
\caption{S(Q,$\omega$) measured at an average scattering angle of $20^\circ$ with an incident energy of 35 meV at four temperatures, normalized on an absolute scale.  The average elastic  momentum transfer is 1.2 \AA$^{-1}$. The solid line is the fit of the crystal field model described in the text with no adjustment of the intensity scale, and the dashed lines represent the individual crystal field transitions. }
\label{Sqw}
\vspace{-0.3in}

\end{center}
\end{figure}

Figure \ref{Sqw} shows low-angle LRMECS data measured with an incident energy of 35 meV at 1.8, 3.2, 5, and 20K.  The data were summed from $2.4^\circ$ to $30^\circ$, giving an average momentum transfer at the elastic position of 1.2 \AA$^{-1}$.   The data in Fig. \ref{Sqw} show that the magnetic spectra are dominated by two crystal field transitions;  the 11 meV peak is a ground-state transition, but the 17.2 meV peak only occurs at higher temperature and represents a transition from a low-lying excited state consistent with the 0.7 meV peak observed on SPINS.  

The data are sufficiently complete to allow an unambiguous determination of the crystal field potential.  The praseodymium ions in \PrOS\  sit on lattice sites with cubic point group symmetry $T_h$, which lacks two symmetry operations of the more common $O_h$ group ($C_4$ and $C^\prime_2$)~\cite{Takegahara:2001}. The crystal field levels in $T_h$ and $O_h$ symmetry have the same degeneracies but different group theory labels and selection rules.  In this article, we will use the more familiar cubic $O_h$ labels, which can be mapped onto the $T_h$ labels using the following: $\Gamma_1(O_h)\rightarrow\Gamma_1(T_h)$, $\Gamma_{3}(O_h)\rightarrow\Gamma_{23}(T_h)$, $\Gamma_4(O_h)\rightarrow\Gamma_4^{(1)}(T_h)$, and $\Gamma_5(O_h)\rightarrow\Gamma_4^{(2)}(T_h)$.  

$T_h$ symmetry requires the addition of an extra crystal field parameter, $B^\prime_6$, to the conventional cubic crystal field Hamiltonian using the Steven's operator formalism:
\begin{equation}
H_{CF} = B_4(O^0_4 + 5O^4_4) + B_6(O^0_6 - 21O^4_6) + B^\prime_6 (O^2_6 - O^6_6)
\end{equation}
The extra parameter, $B^\prime_6$, has a relatively small effect on the energies of the crystal field levels, but  it mixes the $\Gamma_4$ and $\Gamma_5$ wavefunctions so that the usual cubic crystal field selection rules do not apply.  Of particular importance is that the dipole matrix element coupling the $\Gamma_1$ singlet to the $\Gamma_5$ triplet is no longer zero.  These dipole selection rules were a major factor in the previous crystal field assignments, and explain the discrepancy with the conclusions we outline below.

In order to estimate of the importance of this term,  we have performed a Superposition Model analysis of the $B^\prime_6$ parameter for \PrOS~\cite{Newman:1989,Goremychkin:1992}. The starting point of this model is the assumption that the crystal field is the superposition of two-body potentials due to the neighboring ligands.   
\begin{equation}
B_l^m = \Theta_l \sum_i A_l(R_i) K_l^m(\theta_i,\phi_i)
\vspace{-0.1in}
\end{equation}
where $\Theta_l$ are reduced matrix elements , $A_l(R)$ are parameters representing the strength of the two-body potential, and $K_l^m(\theta,\phi)$ are geometric functions tabulated in Ref. \cite{Newman:1989}.   The sum is over all the neighboring ligands at $(R_i, \theta_i, \phi_i)$.  The point-charge model obeys this superposition principle but makes specific predictions concerning the values of $A_l(R)$ that do not affect our conclusions.  If we assume that the crystal field potential is dominated by the nearest-neighbor cage of twelve equidistant antimony ions, the model predicts the ratio of $B^\prime_6$ to $B^0_6$ without any adjustable parameters.  Such a calculation shows that $B^\prime_6$ is substantial in \PrOS. $B^\prime_6/B^0_6=-53.4$, so neglecting this term, as previously proposed~\cite{Tayama:2003}, is not justified.  

In \PrOS, the bulk susceptibility below 5K indicates that the crystal field ground state is non-magnetic.  Therefore, the ground state is either the $\Gamma_1$ singlet or the $\Gamma_{3}$ doublet.  Since the LRMECS data are normalized on an absolute scale and comprise measurements at a number of temperatures, there are four ways in which the two models can be distinguished.

\newcounter{point}
\begin{list}{\alph{point})}{\usecounter{point}}
\item  Assuming that the 0.7 meV transition is the lowest-lying excitation, the absolute cross section of the $\Gamma_3\rightarrow\Gamma_5$ transition is considerably larger than the $\Gamma_1\rightarrow\Gamma_5$ transition and depends sensitively on the value of $B^\prime_6$.  Although the latter transition is not dipole-forbidden in $T_h$ symmetry, it is still relatively weak.  
\item The absolute intensity of the other ground state transition  at 11 meV is very different in the two models.  The $\Gamma_3\rightarrow\Gamma_4$ transition is 25\% stronger than the $\Gamma_1\rightarrow\Gamma_4$ transition.
\item The intensities of the ground state transitions fall much more strongly with a $\Gamma_1$ ground state than with a $\Gamma_3$ ground state because of the greater contrast between the ground state and excited level degeneracies.  For example, increasing the temperature from 1.8~K to 3~K is predicted to reduce the intensity of the 11~meV transition by 15\% for the singlet ground state, but only 5\% for the doublet ground state. 
\item There should be an excited state transition at $\sim$16.5 meV in the case of the $\Gamma_1$ ground state.   No such transition is predicted for a $\Gamma_3$ ground state below 100~K. 
\end{list} 

This gives us confidence that it is possible to establish the crystal field level scheme unambiguously by using all the neutron data in a simultaneous refinement of the crystal field parameters, using the technique discussed in Ref.~\cite{Goremychkin:1992} .  We used the four spectra shown in Fig. \ref{Sqw} and an additional spectrum at 10~K, which is not shown, with the additional constraint of requiring the lowest transition to be at 0.7~meV.  The peak lineshapes were Lorenzians convolved with the instrumental resolution. The linewidths did not vary significantly with temperature and were constrained to be equal in the final refinement. In the least-squares fitting procedure, the only adjustable parameters were the crystal  field potential, the elastic intensity, and the common linewidth.  The reliability of the absolute normalization was such that we did not need to adjust the overall intensity scale. 

\begin{figure}[htb]
\begin{center}

\centerline {
 \includegraphics[width=3in]{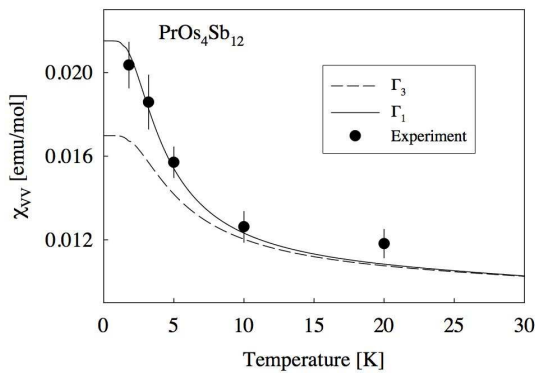}
}
\caption{Temperature dependence of the intensity of the 11 meV peak, represented as a Van Vleck susceptibility (see Ref.~\cite{Goremychkin:1999} for details), compared to the two possible crystal field models. }
\label{ChiVV}
\vspace{-0.3in}

\end{center}
\end{figure}

We performed several refinements with starting parameters consistent with both ground states.  However, it was clear that only the $\Gamma_1$ ground state was consistent with our data.  As Fig. \ref{ChiVV} shows, the temperature dependence of the ground state transitions is too strong to be reproduced accurately with the $\Gamma_3$ ground state.  The intensity of the 0.7~meV peak is too weak in the 6~meV data (not shown).  Furthermore, the $\Gamma_3$ model does not reproduce the excited state transition at 17 meV observed above 10~K.
 
The results of the refinement are shown in Fig. \ref{Sqw}, where all the transitions, from both ground and excited states, are plotted as dashed lines.  It is clear that the $\Gamma_1$ model is able to reproduce all the observed transitions.  The crystal field potential has the following parameters: $B_4  = 0.20(1) \times  10^{-2}$ meV, $B_6 = 0.11(2) \times 10^{-3}$ meV, and $|B^\prime_6| =  0.90(5) \times 10^{-3}$ meV.  The ratio, $B^\prime_6/B_6=8.2$,  is smaller than predicted by the Superposition Model, which can be explained by adding a contribution to the crystal field potential from the osmium sublattice, which contributes to $B_6$, but not to $B^\prime_6$.

The conclusion that the $\Gamma_1$ level is the crystal field ground state in \PrOS\ means that the quadrupolar Kondo effect cannot be responsible for the observed conduction electron mass enhancement.  Nevertheless, it does not rule out other models in which the quadrupolar degrees of freedom of the rare earth f-electrons are important.  As we now discuss, the properties of \PrOS\ can be explained by a delicate balance between two types of interaction, magnetic dipolar and quadrupolar, between the conduction electrons and the praseodymium $f$-electrons.

A theory of conduction electron mass enhancement due to inelastic scattering by crystal field transitions in a singlet ground state system was developed over twenty years ago~\cite{Fulde:1983}. According to Fulde and Jensen~\cite{Fulde:1983}, the mass enhancement due to the inelastic transition at energy $\Delta$ between two levels, labeled $i$ and $j$, is given by

\vspace{-0.1in}
\begin{equation}
\frac{m^*}{m_0} =  1+(g_J-1)^2J_{sf}^2N(0)\frac{2|\langle i|J|j\rangle |^2}{\Delta}
\end{equation}

where $g_J$ is the Land\'e factor, $J_{sf}$ is the exchange integral coupling the conduction electrons to the f-electrons, and $N(0)$ is the bare conduction electron density-of-states at the Fermi level.  If we assume that $N(0)$ is the same as in the isostructural lanthanum compound, then we can use the measured Sommerfeld coefficient in \LaOS, $\gamma=45$~mJ/mol~K$^2$ (averaging over the two published values~\cite{Bauer:2001, Sugawara:2002b}) in order to estimate that $N(0)=3\gamma/2\pi^2k_B^2=9.6\times10^{-3}$~meV$^{-1}$.  We have no reliable estimate for $J_{sf}$, but if we assume that the value derived in praseodymium metal, 0.085~eV~\cite{Fulde:1983},  represents a reasonable order-of-magnitude, we obtain a mass enhancement of $\sim$20.  Given the uncertainty in the value of $J_{sf}$, this is in reasonable agreement with experiment, falling between the estimates based on specific heat ($\sim$50~\cite{Bauer:2002}) and de Haas-van Alphen measurements ($\leq7.6$~\cite{Sugawara:2002b}).

The same $s-f$ exchange that is responsible for the mass enhancement will tend to suppress superconductivity through magnetic pair-breaking~\cite{Abrikosov:1960}.  However, the superconducting transition temperature in \PrOS, T$_c$ = 1.85~K, is 2.5 times larger than the non-magnetic \LaOS,  T$_c$ = 0.74~K~\cite{Sugawara:2002b}.  We propose that the resolution of this apparent discrepancy involves quadrupolar interactions that conserve time-reversal symmetry and therefore enhance pair formation~\cite{Fulde:1970}. This effect, known as aspherical Coulomb scattering, is believed to reduce the rate of suppression of T$_c$ \textit{vs} praseodymium concentration in the singlet ground state system La$_{1-x}$Pr$_x$Sn$_3$~\cite{McCallum:1975, Keller:1976}.  However, this would be the first compound in which T$_c$ is \textit{increased} by praseodymium substitution~\cite{Rotundu:2004b}.  

Fulde \textit{et al} predict that aspherical Coulomb scattering will produce the strongest enhancement of T$_c$ when $\Delta$/T$_c$ is $\sim$10~\cite{Fulde:1970}.  In the case of La$_{1-x}$Pr$_x$Sn$_3$, this ratio only occurs when T$_c$ has already been substantially suppressed by magnetic pair breaking~\cite{Keller:1976}.  However, if we assume that the crystal field potential is nearly constant in the \LPOS\ series, the optimum ratio occurs at $x\rightarrow0$ without requiring any suppression of T$_c$~\cite{Rotundu:2004b}.   The strongest pair-breaking arises from the $\Gamma_1\rightarrow\Gamma_4$ transition, because it has the strongest dipole matrix elements.  This is at much higher energy (11 meV) than the $\Gamma_1\rightarrow\Gamma_5$ transition (0.7 meV), which has a weak dipole but strong quadrupole matrix element, and is therefore responsible for the quadrupolar pair enhancement.  In  La$_{1-x}$Pr$_x$Sn$_3$, these two transitions have comparable energies.  The crystal field level scheme in \PrOS\ is much more favorable for increasing T$_c$ through this mechanism.

In conclusion, we have performed comprehensive inelastic neutron scattering measurements of the temperature dependence of the crystal field transitions in \PrOS\, which strongly suggest that the $\Gamma_1$ singlet is the ground state.  This would rule out the quadrupolar Kondo effect as the mechanism for the heavy fermion state, but favors another scenario in which the observed mass enhancement would arise from inelastic exchange scattering of the conduction electrons by the low-lying crystal field levels.  We argue that inelastic quadrupolar scattering, also known as aspherical Coulomb scattering, provides an explanation for the enhancement in the superconducting transition temperature compared to the isostructural lanthanum compound. The importance of quadrupole interactions in \PrOS\ is evident in the antiferroquadrupolar order observed in high magnetic field~\cite{Ho:2003, Kohgi:2003}.  Our results suggest that it plays a vital role in the superconducting phase as well.

\begin{acknowledgments}
We acknowledge valuable discussions with B. D. Rainford.  This work was performed with the support of the U.S. Department of Energy, Office of Science, under contract no. W-31-109-ENG-38, the U.S. Department of Energy Grant No. FGO2-04ER46105, and the U.S. National Science Foundation Grant No. DMR 0335173.
\end{acknowledgments}

\end{document}
 \end